\RequirePackage{fix-cm}

\documentclass[pdftex,twocolumn,epjc3]{svjour3}  
\RequirePackage[T1]{fontenc}
\RequirePackage{mathptmx}      
\RequirePackage{flushend}
\DeclareMathAlphabet{\mathcal}{OMS}{cmsy}{m}{n} 
\smartqed
\pdfoutput=1 
\usepackage{graphicx,epsfig}
\usepackage{caption}
\usepackage{subcaption}

\usepackage{dcolumn}
\usepackage{bm}
\usepackage{amsmath}
\usepackage{epstopdf}
\usepackage{amsfonts}
\usepackage{amssymb}
\usepackage{xcolor}
\usepackage{float}
\usepackage{multirow}
\usepackage{appendix}
\usepackage{color}
\usepackage{relsize}
\usepackage{xparse}
\usepackage{accents}

\newcommand{\be}{\begin{equation}}
\newcommand{\ee}{\end{equation}}
\newcommand{\bea}{\begin{eqnarray}}
\newcommand{\eea}{\end{eqnarray}}

\newcommand{\abs}[1]{\lvert#1\rvert}

\journalname{Eur. Phys. J. C}
\begin{document}

\title{Reconstructing inflation in scalar-torsion $f(T,\phi)$ gravity
}


\author{Manuel Gonzalez-Espinoza\thanksref{e1,addr1}
        \and Ram\'on Herrera\thanksref{e2,addr1}
        \and
        Giovanni Otalora\thanksref{e3,addr1}
        \and 
        Joel Saavedra\thanksref{e4,addr1}
        }

\thankstext{e1}{e-mail: manuel.gonzalez@pucv.cl}
\thankstext{e2}{e-mail: ramon.herrera@pucv.cl}
\thankstext{e3}{e-mail: giovanni.otalora@pucv.cl}
\thankstext{e4}{e-mail: joel.saavedra@pucv.cl}


\institute{Instituto de F\'{\i}sica, Pontificia Universidad Cat\'olica de
Valpara\'{\i}so, Casilla 4950, Valpara\'{\i}so, Chile.  \label{addr1}
}

\date{Received: date / Accepted: date}

\maketitle

\begin{abstract}
It is investigated the reconstruction during the slow-roll inflation in the most general class of scalar-torsion theories whose Lagrangian density is an arbitrary function $f(T,\phi)$ of the torsion scalar $T$ of teleparallel gravity and the inflaton $\phi$. For the class of theories with Lagrangian density $f(T,\phi)=-M_{pl}^{2} T/2 - G(T) F(\phi) - V(\phi)$, with $G(T)\sim T^{s+1}$ and the power $s$ as constant, we consider a reconstruction scheme for determining both the non-minimal coupling function $F(\phi)$ and the scalar potential $V(\phi)$ through the parametrization (or attractor) of the scalar spectral index $n_{s}(N)$ and the tensor-to-scalar ratio $r(N)$ as functions of the number of $e-$folds $N$. 
 As specific examples, we analyze 
the attractors $n_{s}-1 \propto 1/N$ and $r\propto 1/N$, as well as the case $r\propto 1/N (N+\gamma)$ with $\gamma$ a dimensionless constant. In this sense and depending on the attractors considered, we obtain different expressions for the function $F(\phi)$ and the potential $V(\phi)$, as also the constraints on the parameters present in our model and its reconstruction.
\end{abstract}

\section{Introduction}\label{Introduction}
The Standard Cosmology's problems, such as the horizon, the flatness and the monopole problem, 
can be solved if the early Universe undergoes a period of a quasi-exponential accelerated 
expansion \cite{Guth:1980zm,Starobinsky:1980te,Linde:1981mu}. 
This accelerating phase, known as inflation \cite{Baumann:2009ds,Liddle_Lyth2009,Mukhanov:1990me}, 
is generally accepted as standard at the early Universe. 
Even more, this early inflationary scenario is essential to solve the most puzzling problem of 
standard cosmology, the structure formation problem, due to the generation of primordial density 
fluctuations \cite{Liddle_Lyth2009,Mukhanov:1990me}.  The most successful explanation for the 
physics of inflation is based on a canonical scalar $\phi$ field known as the inflaton, where, the 
simplest case is given by this inflaton minimally coupled to Einstein's gravity with a flat scalar potential 
$V(\phi)$ \cite{MukhanovBook}, and it also can be generalized to the most general second-order scalar-tensor 
theory, the Horndeski gravity theory \cite{Horndeski:1974wa,Kamada:2012se,Kobayashi:2019hrl}. 
\\
\\
On the other hand, the teleparallel equivalent of GR or simply Teleparallel Gravity (TG) is a gauge theory for the translation group which constitutes  an alternative description of gravity based on torsion \cite{Aldrovandi-Pereira-book,JGPereira2,AndradeGuillenPereira-00,Einstein,TranslationEinstein,Early-papers1,Early-papers2,Early-papers3,Early-papers4,Early-papers5,Early-papers6,Arcos:2005ec,Pereira:2019woq}. In this torsion gravity the gravitational field is represented by the translational gauge potential which is the non-trivial part of the tetrad field. So, we can use either the translational gauge potential or the tetrad field as the dynamical variable of theory instead the spacetime metric $g_{\mu \nu}$. The Lorentz connection of TG is the so-called Weitzenb\"{o}ck connection which has torsion but not curvature and then it replaces the torsionless  Levi-Civita connection \cite{Aldrovandi-Pereira-book,JGPereira2,AndradeGuillenPereira-00}. Thus, since the field strength is just the torsion tensor, it is used to construct a torsion-quadratic scalar dubbed torsion scalar $T$ which becomes the Lagrangian density of TG. It differs from the scalar curvature $R$ in a total derivative term and, therefore, the two theories are equivalent in the level of field equations \cite{Aldrovandi-Pereira-book,Arcos:2005ec}. Furthermore, similarly to scalar-tensor theories of gravity \cite{Fujii:2003pa,faraoni2004cosmology}, an interesting extension of TG is a non-minimally coupled scalar-torsion theory \cite{Cai:2015emx,Bahamonde:2017ize}. This kind of torsional modified gravity theories was first considered in Refs. \cite{Geng:2011aj,Geng:2011ka}. It was assumed a non-minimal coupling between the scalar field and torsion in the form $\xi \phi^2 T$, with $\xi$ the coupling constant. It was later extended  in Refs. \cite{Otalora:2013tba,Otalora:2013dsa}, for both, an arbitrary non-minimal coupling function of $\phi$, and a tachyonic kinetic term for the scalar field. An interesting feature of these torsional modified gravity theories is that differently to what happens in scalar-tensor theories where the non-minimal coupling to curvature can be removed through a conformal transformation, in this case the non-minimal coupling to torsion cannot be eliminated under any rescaling of the fields. Although TG coincides with the General Relativity (GR) at the level of field equations these non-minimally coupled scalar-torsion theories belong to a different class of gravitational modifications without any equivalent in curvature-based modified gravity theories \cite{Cai:2015emx}. 
  
The above framework can be further extended by generalizing the action of TG to an arbitrary function $f(T,\phi)$ of the torsion scalar $T$ and a canonical scalar field $\phi$, plus the kinetic term of $\phi$ \cite{Hohmann:2018rwf}. These generalized scalar-torsion $f(T,\phi)$ gravity theories can also be seen as the torsion-based analogue of the so-called generalized $f(R,\phi)$ gravity theories \cite{faraoni2004cosmology}. This general action includes $f(T)$ gravity with scalar field \cite{Yerzhanov:2010vu,Chakrabarti:2017moe,Rezazadeh:2015dza,Goodarzi:2018feh},  non-minimally coupled scalar-torsion gravity \cite{Xu:2012jf,Otalora:2014aoa,Skugoreva:2014ena,Jarv:2015odu,Gonzalez-Espinoza:2019ajd,Jarv:2021ehj}, and its extensions by including a non-linear scalar-torsion coupling \cite{Gonzalez-Espinoza:2020azh}. A non-linear coupling between matter and gravity has been first studied in the context of curvature-based models in Refs. \cite{Nojiri:2004bi,Allemandi:2005qs,Bertolami:2007gv,Harko:2008qz,Harko:2010mv,Bertolami:2013kca}, while the corresponding torsion-based models in Refs. \cite{Harko:2014sja,Carloni:2015lsa,Gonzalez-Espinoza:2018gyl,Harko:2014aja}. Let us note however that a non-minimal coupling to torsion including the kinetic term of the scalar field can lead to a propagation speed of the tensor modes differently from the unit \cite{Gonzalez-Espinoza:2019ajd} which is not compatible with observations \cite{Baker:2017hug,Sakstein:2017xjx}. In the context of inflationary cosmology,  the power spectrum of scalar and tensor perturbations in generalized scalar-torsion $f(T,\phi)$ gravity theories has been recently studied in Ref. \cite{Gonzalez-Espinoza:2020azh}. Here, the authors have shown  that in order to generate primordial fluctuations from generalized scalar-torsion $f(T,\phi)$ gravity, it is necessary to include into the action non-linear terms in torsion like in $f(T)$ gravity plus scalar field or a non-linear coupling to torsion proportional to $F(\phi)G(T)$ where $G(T)$ is a non-linear function of $T$. The case of a linear function $G(T)\sim T$ gives nonzero momentum solutions for the scalaron and then, spoiling the generation of primordial density  fluctuations \cite{Gonzalez-Espinoza:2020azh,Wu:2016dkt}. This kind of non-linear coupling to torsion has also been studied in the context of dark energy in Ref. \cite{Gonzalez-Espinoza:2020jss}, where the authors found new scaling radiation/matter solutions as well as new attractors with accelerated expansion. Finally, in Ref. \cite{Gonzalez-Espinoza:2021mwr} the authors have investigated the stability of scalar perturbations in the presence of a general matter fluid.

In the context of inflation the reconstruction of the background variables (effective potential, coupling functions, scale factor) of the inflationary models, from the observational quantities (the scalar spectrum,  scalar spectral index  and the tensor-to-scalar ratio)  has been studied  by different authors, see Refs. \cite{C,A}. 
In this framework and assuming the slow-roll approximation, a possible methodology for the reconstruction of inflation is the parametrization of these cosmological quantities or parameters in terms of the number of $e-$folds $N$.

In the reconstruction of inflation,  we can consider  the parametrization of the scalar spectral index  $n_{s}(N)$ in terms of the number $N$ or the tensor-to-scalar ratio $r(N)$. Thus,  as an example from these observables and their parametrizations,  we can have  the simple parametrization or attractor $n_{s}(N)= 1-2/N$. This parametrization  for the scalar spectral 
 index $n_{s}(N)$, is well corroborated by Planck satellite \cite{Akrami:2018odb}, 
when the number  $N\simeq$ 60.  Usually,  these observables are evaluated during inflation when the number of  $e-$folds $N\simeq$ 50 $\sim$ 60, and it corresponds to the comoving scale $k$ crossed the Hubble radius i.e.,  when $k=aH$.
 
In the context of the GR and assuming the slow-roll approximation, the attractor  point $n_{s}(N)$ given by $n_{s}(N)= 1-2/N$ gives rise to different effective potentials. In this sense, the reconstruction reproduces the hyperbolic tangent model or T-model \cite{T}, E-model \cite{E}, $R^2$-model \cite{R102} and also the chaotic model in Ref. \cite{R103}. In  the specific case in which we have the reconstruction of inflation with two background  variables as the case of warm inflation,  it is  necessary to assume two attractors; the scalar spectral index $n_{s}(N)$ and the tensor-to-scalar ratio  $r(N)$, in order to rebuild  the effective potential and the dissipation coefficient  as a function of the scalar field, respectively \cite{Herrera:2018cgi}. Analogously, for the reconstruction of Galileon inflation or $G-$inflation with two variables (the effective potential and the coupling parameter in terms of the inflaton field) was necessary to consider the parametrization of the spectral index $n_{s}(N)$ together with the tensor-to-scalar ratio $r(N)$ \cite{Herrera:2018mvo}. Furthermore, the reconstruction of inflation in the slow-roll approximation can be done considering other methodologies. Here, we can mention the parametrization of  the slow-roll parameter $\epsilon(N)$, in terms of the  number of $e-$folds $N$, see e.g. \cite{Huang:2007qz}.  In the same way, the reconstruction of the effective potential and the spectral index from two slow-roll parameters $\epsilon(N)$ and $\eta(N)$, was analyzed in Ref.\cite{Roest:2013fha}. In particular, the reconstruction of the scalar potential, assuming an ansatz in the velocity of the scalar field in terms of the number of $e-$folds, was studied in \cite{Sebastiani:2017cey}. For a review of different reconstruction methodologies in the early scenario, see  Refs. \cite{HH1,HH2}. 

The goal of this research is to reconstruct of a TG inflationary  model, considering the parametrization of the scalar spectral index and the tensor-to-scalar ratio  as a function of the number of $e$-folds. In this framework, we investigate  how the TG  inflationary  model, modifies  the reconstructions of the background variables such as the scalar potential $V(\phi)$ and the coupling function $F(\phi)$. In this context, we will find the structure of the coupling function and  the effective potential, in order to satisfy the  observations imposed by Planck data.

There are previous works related on the reconstruction of $f(T)$ gravity. For example, a reconstruction of $f(T)$ models realizing  inflation at the level of the cosmological background was performed in Ref. \cite{Bamba:2012vg}, by taking some convenient ansatzes for the Hubble parameter and the scale factor. Also, in Ref. \cite{Bamba:2016wjm} the authors have developed a general reconstruction procedure
for $f(T)$ gravity that allows to reproduce a given scale factor evolution, and then they applied it in the inflationary regime. Furthermore, a reconstruction of $f(T)$ gravity with a minimally coupled scalar field was performed in Refs. \cite{Nashed:2014vsa,ElHanafy:2015jbo,ElHanafy:2014efn,Bamba:2016gbu}. The authors applied a special technique by constructing the torsion tensor from a scalar field that allows to construct the effective scalar potential from the adopted (or reconstructed) $f(T)$ gravity theory or inversely to reconstruct the $f(T)$ gravity from some well-known inflationary scalar potential.

On the other hand, in the present work we perform a reconstruction scheme of inflation in the context of the generalized scalar-torsion $f(T,\phi)$ gravity theory. The general action of this theory encompasses the non-minimally coupled scalar-torsion theories as well as $f(T)$ gravity plus scalar field as particular examples. But, the crucial point here is that our reconstruction scheme has as initial input the expressions for the inflationary observables, the spectral index $n_{s}$ and the tensor-to-scalar-ratio $r$. The correct expressions for $n_s$ and $r$ in the context of the generalized scalar-torsion $f(T,\phi)$ gravity theory have been calculated for the first time in Ref. \cite{Gonzalez-Espinoza:2020azh}, where the authors have adequately considered the effects of local Lorentz violation in modified teleparallel gravity. In the present work we use for the first time these expressions for $n_s$ and $r$ for determining both the non-minimal coupling function and the effective scalar potential, through the parametrization (or attractor) of $n_s$ and $r$ as functions of the number of $e$-folds $N$. 

Thus, as an application to the developed formalism, we will analyze  two different  examples on the parametrizations of the observables, in order to rebuild the TG-inflation assuming  the simplest attractor point for the scalar spectral index  $n_s(N)$ and additionally the tensor-to-scalar ratio $r(N)$. In particular we consider the ansatz for the scalar spectral index $n_s(N)$ in both cases  in which  $n_s=1-2/N$ and  for the tensor-to-scalar ratio $r(N)$  we will analyze the cases  $r(N)=q/N$ and $r(N)= q/[N(N+\gamma)]$ where $q$ and $\gamma$ are constants. In this respect, we will reconstruct the effective  potential $V(\phi)$ and the 
non-minimal coupling parameter $F(\phi)$ as a function of the inflaton field $\phi$. Additionally, we will find different constraints on the  parameters in our TG inflationary model from the observational data.

The outline of the paper is as follows: The  Section \ref{TG} we give a brief description  of the scalar-torsion $f(T,\phi)$ gravity. In the Section \ref{GS_torsion_gravity} we discuss the background equations  under the slow-roll approximation and then we review the cosmological perturbations in the frame of  a $f(T,\phi)$ gravity. 
In Section \ref{x4} we find, under a general formalism, explicit relations for the scalar potential and non-minimal coupling function  as a function of the number of $e-$ folds $N$ in order to apply the reconstruction from the observational parameters $n_s(N)$ and $r(N)$. In the Section \ref{HEL}, we study the reconstruction methodology in the high energy limit to get the background variables $V(N)$ and $F(N)$ analytically. Here we consider two specific examples in which we assume the simplest attractor $n_s(N)=1-2/N$ for the scalar spectral index and two ansatzes for the tensor-to-scalar ratio $r(N)$, in order to obtain the effective potential $V(\phi)$ and the non-minimal coupling function $F(\phi)$. Finally, in Section \ref{Concluding_Remarks} we give our conclusions. We chose units so that $c=\hslash=1$.


\section{Scalar-torsion $f(T,\phi)$ Gravity}\label{TG}
Teleparallel Gravity (TG) is a gauge theory for the translation group that constitutes an alternative description of gravity in terms of torsion and not curvature \cite{Aldrovandi-Pereira-book,JGPereira2,AndradeGuillenPereira-00,Einstein,TranslationEinstein,Early-papers1,Early-papers2,Early-papers3,Early-papers4,Early-papers5,Early-papers6,Arcos:2005ec,Pereira:2019woq}. The dynamical variable is the tetrad field $e^{A}_{~\mu}$ and the spacetime metric can be written locally as $g_{\mu \nu}=\eta_{A B} e^{A}_{~\mu} e^{B}_{~\nu}$, where $\eta _{AB}^{}=\text{diag}\,(-1,1,1,1)$ is the Minkowski tangent space metric \cite{Aldrovandi-Pereira-book,Pereira:2019woq}. Similarly as in curvature-based modified gravity models  \cite{Fujii:2003pa,faraoni2004cosmology,Tsujikawa:2008uc,Alimohammadi:2009yt}, we can modify gravity starting from the action of teleparallel gravity \cite{Aldrovandi-Pereira-book,JGPereira2,AndradeGuillenPereira-00} by extending it to the general scalar-torsion gravity action \cite{Hohmann:2018rwf,Gonzalez-Espinoza:2020azh}
\begin{equation}
 S=\int d^{4}x\,e\,\left[ f(T,\phi)+ P(\phi)X \right],
\label{action1}
\end{equation}
where $e=\det{(e^{A}_{~\mu})}=\sqrt{-g}$, and $f$ is an arbitrary function of the torsion scalar and of the  scalar field $\phi$.
 The torsion scalar $T$ is given by
 \be
T=\frac{1}{4} T^{\rho \mu \nu} T_{\rho \mu \nu} +\frac{1}{2} T^{\rho \mu \nu} T_{\nu \mu \rho}-T_{\rho \mu}^{~~~\rho} T^{\nu \mu}_{~~~~\nu},
\label{TorsionScalar}
 \ee which is quadratic in the torsion tensor 
 \be
 T^{\rho}_{~\mu \nu}=e_{A}^{~\rho}\left(\partial_{\mu}{e^{A}_{~\nu}}-\partial_{\nu}{e^{A}_{~\mu}}+\omega^{A}_{~B \mu} e^{B}_{~\nu}-\omega^{A}_{~B \nu} e^{B}_{~\mu}\right),
 \label{TorsionTensor}
 \ee that is associated to the Weitzenb\"{o}ck connection of teleparallel gravity $T^{\rho}_{~\nu \mu}=\Gamma^{\rho}_{~~\mu \nu}-\Gamma^{\rho}_{~~\nu \mu}$ and $\omega^{A}_{~B \mu}$ is the spin connection \cite{Aldrovandi-Pereira-book,JGPereira2,AndradeGuillenPereira-00}.  The kinetic term is the product of the arbitrary function $P(\phi)$ and $X=-\partial_ {\mu}{\phi}\partial^{\mu}{\phi}/2$. This general action encompasses a large class of torsion-based modifications of gravity, such as $f(T)$ gravity, plus minimally coupled scalar field, and non-minimally coupled scalar-torsion gravity models. For the function $f(T,\phi)=-M_{pl}^2 T/2-V(\phi)$, we recover TG plus scalar field, with $V(\phi)$ the scalar potential \cite{MukhanovBook}. 
 Here $M_{pl}$ denotes the reduced Planck mass and it corresponds to $M_{pl}=1/(8\pi \mathcal{G})^{1/2}$ with $\mathcal{G}$ the gravitational constant.
Although TG is equivalent to GR at level of field equations, 
the action \eqref{action1} constitutes a different class of gravitational modifications with rich phenomenology without any equivalent in the context of the curvature-based gravitational modifications \cite{Cai:2015emx}.

Varying with respect to the tetrad field $e^{A}_{~\mu}$,  we obtain the modified field equations
\bea
 && f_{,T} G_{\mu \nu}+S_{\mu \nu}{}^{\rho} \partial_{\rho} f_{,T}+\frac{1}{4}g_{\mu \nu}\left(f-T f_{,T}\right)+\nonumber\\ && \frac{P}{4}\left(g_{\mu \nu} X+\partial_{\mu}\phi \partial_{\nu}\phi\right)=0,
\label{FieldEquations}
\eea where $S_{\mu\nu}{}^{\rho}$ is the dubbed superpotential which is linear in the torsion tensor \cite{Aldrovandi-Pereira-book}. Also, in this equation $G^{\mu}_{~\nu}=e_{A}^{~\mu} G^{A}_{~\nu}$ is the Einstein tensor with $G_{A}^{~\mu}\equiv e^{-1}\partial_{\nu}\left(e e_{A}^{~\sigma} S_{\sigma}^{~\mu\nu}\right)-e_{A}^{~\sigma} T^{\lambda}_{~\rho \sigma}S_{\lambda}^{~\rho \mu}+e_{B}^{~\lambda} S_{\lambda}^{~\rho \mu}\omega^{B}_{~A \rho}+\frac{1}{4}e_{A}^{~\mu} T$ \cite{Aldrovandi-Pereira-book}. In the equation \eqref{FieldEquations}, we have expressed the field equations in a general coordinate basis by contracting with the tetrad field. Since the action \eqref{action1} is not local Lorentz invariant the field equations \eqref{FieldEquations} are not symmetric. In fact, the superpotential tensor $S_{\mu \nu}{}^{\rho}$  is not symmetric in the lower indices. Then, at the moment of perturbing the cosmological background is necessary to take into account the additional degrees of freedom associated with the violation of local Lorentz invariance \cite{Gonzalez-Espinoza:2020azh}.

On the other hand, by varying with respect to the scalar field we obtain the motion equation
\be
\bar{\nabla}_{\mu}\left(P \partial^{\mu}{\phi}\right)+P_{,\phi} X+f_{,\phi}=0,
\label{MotionEq}
\ee  where we have denoted $f_{,\phi}=\partial{f}/\partial \phi$ and $P_{,\phi}=d P/d\phi$. Also, we have introduced the covariant derivative $\bar{\nabla}_{\mu}$ of the Levi-Civita Connection $\bar{\Gamma}^{\rho}_{~\mu \nu}$ which is related to the Weitzenb\"{o}ck connection $\Gamma^{\rho}_{~\mu \nu}$ through the relation $\Gamma^{\rho}_{~\mu \nu}=\bar{\Gamma}^{\rho}_{~\mu \nu}+K^{\rho}_{~\mu \nu}$ with $K^{\rho}_{~\mu \nu}$ the contortion tensor \cite{Aldrovandi-Pereira-book}.

\section{Slow-roll inflation in scalar-torsion $f(T,\phi)$ Gravity}\label{GS_torsion_gravity}

 In this section  we give a brief description of the background equations in 
the context of 
the slow-roll approximation and   we also review  the cosmological perturbations in this gravity. 
In this sense, we start  with
the standard homogeneous and isotropic background geometry by choosing
\begin{equation}
\label{veirbFRW}
e^A_{~\mu}={\rm
diag}(1,a,a,a),
\end{equation}
that is the proper tetrad naturally associated with the vanishing spin connections $\omega^{A}_{~ B\mu}=0$ \cite{Krssak:2015oua}, and it gives  the flat Friedmann-Robertson-Walker
(FRW) metric 
\begin{equation}
ds^2=-dt^2+a^2\,\delta_{ij} dx^i dx^j \,,
\label{FRWMetric}
\end{equation}
where $a$ is the scale factor which is a function of the cosmic time $t$. 

Replacing this tetrad field in the field equations \eqref{FieldEquations} and \eqref{MotionEq} we obtain the background equations
\bea
\label{00}
 f(T,\phi) - P(\phi) X - 2 T f_{,T}=0, \\
\label{ii}
 f(T,\phi) + P(\phi) X - 2 T f_{,T} - 4 \dot{H} f_{,T} - 4 H \dot{f}_{,T}=0, \\
\label{phi}
 P(\phi) \ddot{\phi} + 3 P(\phi) H \dot{\phi}+ P_{,\phi} X- f_{,\phi}=0,
\eea
where $H\equiv \dot{a}/a$  corresponds to the Hubble rate. In the following will assume that
a dot 
represents derivative with respect to $t$,
{and the notation in which }
a comma denotes derivative with respect to $\phi$ or $T$  i.e., $f_{,T}$ corresponds to 
$\partial f/\partial\,T$, $f_{,\phi}$ to $\partial f/\partial\phi$,  etc. 
Also, we have used the relation between the torsion scalar and the Hubble parameter, $T=6 H^2$, that is obtained after substituting  Eq. \eqref{veirbFRW} into Eq. \eqref{TorsionTensor} and then into Eq. \eqref{TorsionScalar}.

 Now we introduce the slow-roll parameters  that are defined as
\bea
&& \epsilon= -\dfrac{\dot{H}}{H^2},\:\:\: \delta_{P X} = - \dfrac{P(\phi) X}{2 H^{2} f_{,T}},\:\:\ \delta_
{f_{,T}} = \dfrac{\dot{f}_{,T}}{f_{,T} H},\nonumber\\
&&  \delta_{P}=\frac{\dot{P}}{H P},\:\:\:\: \delta_{\phi}=\frac{\ddot{\phi}}{H\dot{\phi}}\,\,\,.
\label{slowpara1}
\eea Thus, from Eqs. \eqref{00} and \eqref{ii} we obtain
\be
\epsilon=\delta_{PX} +\delta_{f_{,T}}.
\label{slow_roll_eps}
\ee Also, it is useful to define
\be
\delta_{f\dot{H}}=\frac{f_{,TT} \dot{T}}{H f_{,T}},\:\:\:\: \delta_{fX}=\frac{f_{,T\phi}\dot{\phi}}{H f_{,T}},
\ee such that
\be
\delta_{f_{,T}}=\delta_{f\dot{H}}+\delta_{fX}.
\ee
Under the slow-roll approximation,  in which
$\epsilon$, $\delta_{PX}$, $\delta_{f_{,T}}$, $\delta_{P}$, $\delta_{\phi} \ll 1$,   during the inflationary epoch, 
then 
the background equations can be  reduced to 
\bea
 f(T,\phi) &\simeq& 2 T f_{,T},\label{bg_sr1} \\
3 P(\phi) H \dot{\phi} &\simeq&   f_{,\phi}\label{bg_sr2}.
\eea
In order to study the reconstruction for scalar-torsion $f(T,\phi)$ gravity, 
 we will analyze the particular case in which the 
Lagrangian density  corresponds to  \cite{Gonzalez-Espinoza:2020azh}
\be
f(T,\phi)=-\dfrac{M_{pl}^{2}}{2} T - G(T) F(\phi) - V(\phi),\label{Ayy}
\ee 
 where we have considered for simplicity that the function $P(\phi)=1$. Here, the quantity 
$G(T)$ is a coupling 
 function   that depends exclusively on the torsion scalar $T$ and we can 
 identify this function as the gravitational coupling associated to the torsion scalar. In the context of inflation, different 
 functions
 $G(T)$  have been studied in the literature, see e.g., 
 Refs.\cite{Geng:2011aj,Gonzalez-Espinoza:2020azh,Jamil:2012vb}.

 Now using equation \eqref{bg_sr1} we have 

\begin{equation}
\dfrac{M^2_{pl}}{2} T \simeq  V - G(T) F(\phi) C(T),
\label{T_phi}
\end{equation}
where   the quantity $C(T)$ is defined as 
$C(T)= T \dfrac{\partial}{\partial T} \text{ln} \left( \dfrac{G^2}{T} \right) $.  We mention that  in the specific 
case in which the function $G(T) \propto T^{s}$ with $s=$ constant, we obtain that the function $C(T)$ 
corresponds to a constant.

 We note that the Eq. (\ref{T_phi}) can be written  in the standard form as 
$3H^2=8\pi\, \mathcal{G}_{eff}\,V$, with which 
 we can identify that  the effective gravitational coupling $\mathcal{G}_{eff}$  is given 
 by
\be
\mathcal{G}_{eff}(T,\phi)= \mathcal{G}_{eff}=\frac{\mathcal{G}}{1+\frac{2 C(T) F(\phi) G(T)}{M_{pl}^2 T}}. 
\label{Gral_Geff}
\ee 

Also, from Eq. \eqref{bg_sr2} one finds
\bea
 \frac{\dot{\phi}}{M_{pl} H} &\simeq& -\Bigg[\frac{2 G F_{,\phi}}{M_{pl} T}+\frac{2 V_{,\phi}}{M_{pl} T}\Bigg].
\label{dotphioverH}
\eea 
 From the slow-roll equation (\ref{dotphioverH}), we may distinguish two 
situations for the velocity of the scalar field; when $\dot{\phi}<0$ or 
$\dot{\phi}>0$. For example, when the functions  
$G>0$, 
$F_{,\phi}>0$ and $V_{,\phi}>0$, then the velocity associated to scalar field 
is negative i.e.,
 $\dot{\phi}<0$. Inversely, if $G>0$, $F_\phi<0$ and  $V_\phi<0$ then $\dot{\phi}>0$,  and so on 
 other combinations of the functions; such that $\dot{\phi}\lessgtr 0$ from (\ref{dotphioverH}).
  In this context, in the following 
 we will assume that the velocity associated to scalar field is $\dot{\phi}<0$.  

From the definition  given by Eq. \eqref{slowpara1},  we can rewrite  the slow-roll
 parameters $\delta_{PX}$ and $ 
\delta_{f,T}$ as 
\bea
\delta_{PX}=\frac{1}{2\left(1+\frac{2 G_{,T} F}{M_{pl}^2}\right)}\left(\frac{\dot{\phi}}{M_{pl} H}\right)^2,
\eea and

\bea
\delta_{f,T}&=&\Bigg[ \left(\frac{2 G_{,T} F_{,\phi}}{M_{pl}^3}\right)
\left(\frac{\dot{\phi}}{M_{pl} H}\right)-\nonumber\\
&& \left(\frac{2 T G_{,TT} F}{M_{pl}^4\left(1+\frac{2 G_{,T} F}{M_{pl}^2}\right)}\right)\left
(\frac{\dot{\phi}}{M_{pl} H}\right)^2\Bigg]\times\nonumber\\
&& \left(1+\frac{2 G_{,T} F}{M_{pl}^2}+\frac{4 T G_{,TT} F}{M_{pl}^2}\right)^{-1}.
\eea

Thus, by using the relation \eqref{slow_roll_eps}, and \eqref{dotphioverH}, one obtains
\bea
\epsilon&=&\frac{\left(\frac{\dot{\phi}}{M_{pl} H}\right)^2+\frac{4 F_{,\phi} G_{,T}}{M_{pl}}
 \left(\frac{\dot{\phi}}{M_{pl} H}\right)}{2\left(1+\frac{2 G_{,T} F}{M_{pl}^2}+\frac{4 T G_{,TT} F}
 {M_{pl}^2}\right)}\nonumber\\
 &\simeq& \frac{2 V_{,\phi}^2}{M_{pl}^2 T^2 \left(1+\frac{2 G_{,T} F}{M_{pl}^2}+
\frac{4 T G_{,TT} F}{M_{pl}^2}\right)}+\nonumber\\
&& \frac{4 F_{,\phi} V_{,\phi} \left(G-T G_{,T}\right)}{M_{pl}^2 T^2 \left(1+\frac{2 G_{,T} F}{M_{pl}^2}+
\frac{4 T G_{,TT} F}{M_{pl}^2}\right)}+\nonumber\\
&&\frac{ 2 F_{,\phi}^2 G \left(G-2 T G_{,T}\right)}{M_{pl}^2 T^2 \left(1+\frac{2 G_{,T} F}{M_{pl}^2}+
\frac{4 T G_{,TT} F}{M_{pl}^2}\right)},
\label{Full_epsilon}
\eea
and then also
\bea
 \delta_{\phi} &\simeq& \epsilon-\frac{4 F_{,\phi}^2 G_{,T} \left(G-2 T G_{,T}\right)}{M_{pl}^2 T
 \left(1+\frac{2 G_{,T} F}{M_{pl}^2}+\frac{4 T G_{,TT} F}{M_{pl}^2}\right)} -\nonumber\\
&& \frac{4 F_{,\phi} V_{,\phi} G_{,T}}{M_{pl}^2 T \left(1+\frac{2 G_{,T} F}{M_{pl}^2}+\frac{4 T G_{,TT} F}{M_{pl}^2}
\right)}-\nonumber\\
&& \frac{2 F_{,\phi\phi} G}{ T}-\frac{2 V_{,\phi\phi}}{ T}.
\eea

On the other hand, the scalar and tensor power spectral  of primordial fluctuations for the scalar-torsion 
$f(T,\phi)$ gravity were obtained in Ref. \cite{Gonzalez-Espinoza:2020azh}. The scalar power spectrum 
of curvature perturbation is given by
\bea
 \mathcal{P}_{s}(k)&\equiv& \frac{k^3}{2 \pi^2}\left|\mathcal{R}_{k}(\tau)\right|^2 \simeq 
\frac{H_{k}^2}{8 \pi^2 Q_{sk}}\left[1+2\eta_{\mathcal{R}}\ln\left(\frac{k}{a H}\right)\right],\nonumber\\
&\simeq& \frac{T}{96 \pi^2 M_{pl}^2 \left[\frac{G F_{,\phi}}{M_{pl} T}+
\frac{V_{,\phi}}{M_{pl} T}\right]^2}.\label{mm}
\eea  In the first line of this latter equation, we have $Q_{s k}=\left.\frac{P X}{H^2}\right|_{k=a H}$ and 
$H_{k}=\left.H\right|_{k=a H}$, where these quantities are evaluated when the wavelength of the 
perturbation crosses the Hubble radius, i.e., $k=aH$.

Following  Ref. \cite{Gonzalez-Espinoza:2020azh}, the mass term  associated to the parameter $\eta_{\mathcal{R}}$  (see eq.(\ref{mm})) can be 
 written in terms of the 
slow-roll parameters  such that
\bea
&&\eta_{\mathcal{R}}=\frac{m^2}{3 H^2}=\delta_{f_{,T}}\left[1 + \left(1+\frac{\delta_{fX}}{\delta_
{PX}}\right)\dfrac{\delta_{f_{,T}}}{\delta_{f\dot{H}}} \right],\nonumber\\
\label{mass_term} 
\eea 
 and  during the inflationary scenario 
$\abs{\eta_{\mathcal{R}}}\ll 1$, see \cite{Gonzalez-Espinoza:2020azh}. 

 As the scalar spectral index $n_s$ associated with the spectrum $\mathcal{P}_{s}(k)$ is given by 
$n_{s}-1\equiv\frac{d \ln{\mathcal{P}_{s}(k)}}{d\ln{k}}$,  we get
\bea
n_{s}-1 &\equiv& \left.\frac{d \ln{\mathcal{P}_{s}(k)}}{d\ln{k}}\right|_{k=a H}=-4\epsilon-
\delta_{P}-2\delta_{\phi}+2 \eta_{\mathcal{R}}\nonumber\\
&\simeq& -\frac{4 V_{,\phi}^2 }{M_{pl}^2 T^2}\left[\frac{2}{1+\frac{2 F G_{,T}}
{M_{pl}^2}}+\frac{1}{1+\frac{2 F \left(2 T G_{,TT}+G_{,T}\right)}{M_{pl}^2}}\right]-\nonumber\\
&& \frac{8 F_{,\phi} V_{,\phi}}{M_{pl}^2 T^2} \left[\frac{2 G}{1+\frac{2 F G_{,T}}
{M_{pl}^2}}-\frac{T G_{,T}-G}{1+\frac{2 F \left(2 T G_{,TT}+G_{,T}\right)}{M_{pl}^2}}\right]-\nonumber\\
&& \frac{4 F_{,\phi}^2}{M_{pl}^2 T^2} \Bigg[\frac{2 G^2}{1+\frac{2 F G_{,T}}
{M_{pl}^2}}-\frac{G\left(2 T G_{,T}-G\right)}{1+\frac{2 F \left(2 T G_{,TT}+G_{,T}\right)}{M_{pl}^2}}+\nonumber\\
&& \frac{ M_{pl}^2 T G_{,T}^2}{F G_{,TT}}\Bigg]+\frac{4 F_{,\phi\phi} G}{T}+\frac{4 V_{,\phi\phi}}{ T}.
\label{nS}
\eea
Also, the tensor power spectrum becomes
\be
\mathcal{P}_{T}=\frac{H_{k}^2}{2 \pi^2 Q_{Tk}}, 
\ee where $Q_{Tk}=-\left.\frac{f_{,T}}{2}\right|_{k=a H}$, and then, the tensor-to-scalar ratio takes the form
\bea
 r = \dfrac{\mathcal{P}_{T}}{\mathcal{P}_{s}} \simeq 16 \delta_{PX}\simeq
  \frac{32 \left[G F_{,\phi}+ V_{,\phi}\right]^2}{M_{pl}^2 T^2 
  \left(1+\frac{2 F G_{,T}}{M_{pl}^2}\right)}.
\label{r_phi}
\eea

 In the following we will analyze the reconstruction of our model from the observable 
parameters;
the scalar spectral index and tensor-to-scalar ratio in terms of the number of $e-$ folding $N$.

\section{Reconstruction from the attractors $n_{s}(N)$ and $r(N)$}\label{x4}

 In this section, we will  apply  the methodology  considered  to
reconstruct the background variables (the effective potential $V(\phi)$ and the  non-minimal coupling  function $F(\phi)$), assuming as attractors the scalar spectral index and the tensor-to-scalar ratio as functions of the number of $e-$folds $N$.

In the framework of this reconstruction, we rewrite the background variables, together with the  spectral index and the tensor-to-scalar ratio as  functions of the number of $e$-folds $N$. Under these relations and assuming the scalar spectral index  $n_{s}=n_{s}(N)$ together with the tensor-to-scalar ratio $r(N)$, we should find the scalar potential $V$ and the non-minimal coupling parameter $F$ as functions of the number of $e-$folds $N$.

Thus, 
 to give
 a measure of the inflationary expansion during inflation, we can  define
 the number of $e-$folds $N$
 from a particular time $t$ until the end of inflation at the 
time $t_{f}$ wherewith

\bea
&& N=\log(a_{f}/a)= \int_{t}^{t_{f}} H \text{d} t=\int_{\phi}^{\phi_{f}}\frac{H}{\dot{\phi}}d\phi.
\label{N_efolds}
\eea 
Thus, from this relation
  we should
  obtain analytically  the number of $e-$folds $N$ in terms of the inflaton field $\phi $ 
  i.e., $N=N(\phi)$
and then 
 we should  reconstruct  the scalar potential $V(\phi)$ and the coupling function $F(\phi)$. 
 Here we have considered that the number of $e-$folds at the end of 
 the inflationary epoch is defined as $N(t=t_f)=0$.

So, by using equation  \eqref{dotphioverH}, one finds 
\bea
&& \frac{dN}{d\phi}=\phi_{,N}^{-1}
=\frac{1}{M_{pl}}\left[\frac{2 G F_{,\phi}}{T M_{pl}}+\frac
{2 V_{,\phi}}{M_{pl} T}\right]^{-1}>0,
\label{dNdphi}
\eea 
 since $\dot{\phi}=-\phi_{,N} H$. We mention that the function $G(T)$ can be written as a function of $\phi$ once that the torsion scalar is solved 
as a function $T=T(\phi)$ from Eq. \eqref{T_phi}. 
Also, we have
\be 
V_{,\phi}=\frac{V_{,N}}{ \phi_{,N}},\:\:\:\: F_{,\phi}=\frac{F_{,N}}{ \phi_{,N}},
\label{V_phi}
\ee  and 
\be
V_{,\phi\phi}=\frac{V_{,NN}}{\phi_{,N}^2}-\frac{V_{,N}\phi_{,NN}}{\phi_{,N}^{3}},\:\:\:F_{,\phi\phi}=
\frac{F_{,NN}}{\phi_{,N}^2}-\frac{F_{,N}\phi_{,NN}}{\phi_{,N}^{3}}.
\label{V_phiphi}
\ee

Putting the relations \eqref{V_phi} in Eq.\eqref{dNdphi}, 
we find  that the relation between the scalar field and the number $N$ can be written as
\be
\phi_{,N}=\sqrt{\frac{2}{T}\left(F_{,N} G+V_{,N}\right)}\,\,,
\label{phi_N2}
\ee  or equivalently, 
\be
\int{\sqrt{\frac{2}{T}\left(F_{,N} G+V_{,N}\right)}dN}=\int{d\phi},
\label{N_phi}
\ee  
where the combination $F_{,N}\;G+V_{,N}>0$. 
 From this relation, we note that it is possible  to obtain  the number of the $e-$ folds in terms of the scalar 
field i.e.,  
$N=N(\phi)$. We mention that in the methodology  of the reconstruction is important to obtain the 
relation $N=N(\phi)$ and it is 
 only possible   in the situation in which we can invert the solution given by Eq. (\ref{N_phi}).

 Also, differentiating this equation one has 
\bea
\phi_{,NN}&=&-\frac{T_{,N} \left(F_{,N} \left(G-T G_{,T}\right)+V_{,N}\right)}{\sqrt{2 T^3
\left(F_{,N} G+V_{,N}\right)}}+
\nonumber\\
&&
\frac{F_{,NN} G+V_{,NN}}{\sqrt{2 T\left(F_{,N} G+V_{,N}\right)}},
\eea 
 and then by using these expressions back in Eqs. \eqref{V_phi} and \eqref{V_phiphi}, one obtains
\bea
V_{,\phi}&=&\sqrt{\frac{T}{2 \left(F_{,N} G+V_{,N}\right)}} V_{,N},\\
V_{,\phi\phi}&=& \frac{T_{,N} V_{,N} \left(V_{,N}+F_{,N} G-T F_{,N} G_{,T}\right)}{4 \left(F_{,N} G+
V_{,N}\right)^2}+
\nonumber\\
&& 
\frac{T V_{,NN} \left(2 F_{,N} G+V_{,N}\right)-T F_{,NN} G V_{,N}}{4 \left(F_{,N} G+V_{,N}\right)^2}.
\eea  
Similarly,  for the function $F$ we have
\bea
F_{,\phi}&=&\sqrt{\frac{T}{2 \left(F_{,N} G+V_{,N}\right)}} F_{,N},\\
F_{,\phi\phi}&=&\frac{F_{,N} T_{,N} \left(V_{,N}+F_{,N} G-T F_{,N} G_{,T}\right)}{4 \left(F_{,N} G+
V_{,N}\right)^2}+
\nonumber\\
&& 
\frac{T F_{,NN} \left(F_{,N} G+2 V_{,N}\right)-T F_{,N} V_{,NN}}{4 \left(F_{,N} G+V_{,N}\right)^2}.
\eea
Therefore, substituting the above expressions  in Eqs. \eqref{nS} and \eqref{r_phi},  we can rewrite the scalar spectrum index 
and the tensor-to-scalar ratio as 
\bea
\label{ns_N}
n_{s}(N)-1 &=& T_{,N} \left[\frac{1}{T}-\frac{F_{,N} G_{,T}}{F_{,N} G+V_{,N}}\right]+\frac{V_{,NN}}{F_{,N} G+V_{,N}}+
\nonumber\\
&& 
\frac{F_{,NN} G}{F_{,N} G+V_{,N}}-\frac{4 \left(F_{,N} G+V_{,N}\right)}{ T\left(M_{pl}^2+2 F G_{,T}\right)}-\nonumber\\\
&& \frac{2 F_{,N} \left(G-2 T G_{,T}\right)+V_{,N}}{T \left(M_{pl}^2+2 F \left(2 T G_{,TT}+G_{,T}\right)\right)}\nonumber\\
&&-\frac{2 F_{,N}^2 G_{,T}^2}{F G_{,TT} \left(F_{,N} G+V_{,N}\right)},\\
\mbox{and}\,\,\,\,\,\,\,\,\,\,\,\,\,\,\,\,\,\,\,\,\,\,\,&&  \nonumber\\
r(N)&=&\frac{16 \left(F_{,N} G+V_{,N}\right)}{T \left(M_{pl}^2+2 
F G_{,T}\right)}.
\label{r_N}
\eea 
  Here, the quantities $T(N)$ and therefore $G(T(N))=G(N)$ are functions of 
 $N$.

  Thus, Eqs. \eqref{N_phi}, \eqref{ns_N}, and \eqref{r_N} constitute the set of basic 
  equations for reconstructing the potential scalar $V(\phi)$ and the non-minimal coupling 
  function $F(\phi)$.  In the following, we will analyze some particular examples in order to rebuild these
  background variables from the parametrization of the  scalar spectral index and the tensor-to-scalar ratio
  as functions of the number of $e-$folds $N$. Additionally,  in the following we will consider the high 
  energy limit in which the torsion $T\ll \mid G(T)F(\phi)C(T)\mid/M_{pl}^2\sim V/M_{pl}^2$, in order to find   analytical 
  expressions  for the effective potential and non-minimal coupling in terms of the scalar field from 
  the reconstruction. 

\section{High Energy Limit}\label{HEL}

 In order to consider the formalism of above and to reconstruct analytically 
the effective potential $V$ and the non-minimal coupling $F$, we shall take the the high 
  energy limit in which the scalar torsion $T\ll \mid G(T)F(\phi)C(T)\mid/M_{pl}^2\sim V/M_{pl}^2$. 
Also, in the following we will consider a power-law dependence on the scalar torsion for the coupling 
function $G(T)$ given by  
 $G(T)=T^{s+1}$, where the power $s$ corresponds to a constant and it is a real quantity. This power-law 
 dependence for the function $G(T)$ has been considered in Refs. \cite{Geng:2011aj,Jamil:2012vb} for models of 
 dark energy in the framework of TG and for the inflationary model in \cite{Gonzalez-Espinoza:2020azh}.

Thus, under these considerations  the Eq. \eqref{T_phi}, is reduced to 
\be
\frac{M_{pl}^2}{2}T+(2 s+1) F T^{s+1}\simeq(2 s+1) F T^{s+1}=V,
\label{TEq1}
\ee 
and  the effective gravitational constant becomes
\be
\mathcal{G}_{eff}=\frac{\mathcal{G}}{1+\frac{2\left(2s+1\right) F T^{s}}{M_{pl}^2}},
\ee 
 where we have that
in the strong coupling limit the rate  $\mathcal{G}_{eff}/\mathcal{G} \ll 1$.

 In this way, from Eq. \eqref{TEq1} one finds
\be
T\simeq \left[\frac{V}{(2 s+1) F}\right]^{1/(s+1)}.
\label{Strong_T}
\ee 

We note that  assuming  the coupling function $F$ as positive, then the  
power $s$ associated to the function $G(T)$ is  $s>-1/2$.

Now, combining Eqs. \eqref{phi_N2} and \eqref{Strong_T}, we find that 
 the quantity $\phi_{,N}$ 
can be written as 
\be
\phi_{,N}\simeq\sqrt{2}  (2 s+1)^{\frac{1}{2 (s+1)}}\left(\frac{F}{V}\right)^{\frac{1}{2 (s+1)}}\sqrt{\frac{V F_{,N}}{(2 s+1) F}+V_{,N}}.
\label{phi_N_Strong}
\ee 
In this form, under the high energy limit, we obtain that 
the scalar spectral index $n_{s}$ and the tensor-to-scalar ratio $r$ in terms 
of the number of $e-$folds from Eqs. \eqref{ns_N} and  \eqref{r_N} are reduced to
\bea
\label{ns_phi_Strong}
n_{s}(N)-1 &\simeq& -\frac{2 (2 s+1) V_{,N}}{ (s+1)V}-\frac{(5 s+3) F_{,N}}{ (s+1) (2 s+1) F}-
\nonumber\\
&&
\frac{2 (s+1)^2 V F_{,N}^2}{s (2 s+1) F \left[V F_{,N}+(1+2 s) F V_{,N}\right]}+\nonumber\\
&&\frac{(2 s+1) F V_{,NN}}{V F_{,N}+(2 s+1) F V_{,N}}+\nonumber\\
&&\frac{V F_{,NN}}{V F_{,N}+(2 s+1) F V_{,N}},
\eea and
\bea
&& r(N)\simeq \frac{8 F_{,N}}{(s+1) F}+\frac{8 (2 s+1) V_{,N}}{(s+1) V},
\label{r_phi_Strong}
\eea 
respectively.

Combining the equations \eqref{ns_phi_Strong} and \eqref{r_phi_Strong}, along 
with the derivatives of Eq. \eqref{r_phi_Strong}, we can obtain a decoupled system 
with first-order equations
for the coupling function $F$ and the effective potential $V$ as functions of the number of $e-$folds
 given by
\bea 
\label{F_N_Strong}
&& F_{,N}= \frac{1}{32} s F \left(-3 r\pm \sqrt{A}\right),\\
&& V_{,N}= \frac{ V \left[(7 s+4) r\mp s\sqrt{A}  \right]}{32 (2 s+1)}
\label{V_N_Strong},
\eea 
where we have defined the function $A(n_s,r)=A(N)=A>0$ as 

\bea
A&=&\frac{r\left[64 (2 s+1) (1-n_{s})-(15 s+8) r\right]}{s}+\nonumber\\
&& \frac{64 (2 s+1) r_{,N}}{s}.
\eea

As the function $A$ is positive, then we find the constraints
\be
r_{,N}\lessgtr r\left[\frac{\left(15 s+8\right)r}{64(2 s+1)}-(1-n_{s})\right],\label{rcN}
\ee 
where the symbol $<$ corresponds to the values of $s$ between $-1/2<s<0$  and the symbol $>$ for the situation in which  $s>0$. In particular for the specific case in which the constant $s\gg 1$ (recalled that the parameter $s>-1/2$), then the lower limit of the derivative $r_{,N}$ becomes $r[15 r/128 -(1-n_s)]<r_{,N}$. In the opposite case, when $0<s\ll 1$,  we have that the lower limit
for $r_{,N}$ is given by $r[(1-s/8)(r/8)-(1-n_s)]<r_{,N}$

Now, from Eqs. \eqref{F_N_Strong} and \eqref{V_N_Strong} we
find that the function $F$ and the scalar potential $V$ in terms of the number of $e-$ 
folds $N$ can be written in   general as
\bea 
\label{F_N_Strong2}
&& F(N)=F_0\, \exp\left[\int_{N_0}^N\,\frac{s\left(-3 r\pm \sqrt{A}\right)}{32}\,dN'\right],\\
&& V(N)=V_0\exp\left[\int_{N_0}^N\,\frac{(7 s+4) r\mp s\sqrt{A}}{32 (1+2 s)}\,\,dN'\right],
\label{V_N_Strong2}
\eea 
where $V_0>0$ and $F_0>0$ are two integration constants
defined as $F(N=N_0)=F_0$ and $V(N=N_0)=V_0$, in which  $N_0$  denotes the number of $e-$folds during the slow-roll scenario and its value is such that $N>N_0>0$.

From these equations and  considering a suitable ansatz for the observable parameters $n_s(N)$ and also for $r(N)$, we shall obtain the background variables $F$ and $V$ in terms of the number of $e-$folds, in order  to continue with the reconstruction of our model.

In the following, during the high energy regime we will consider some
examples for the parametrization on the observable parameters $n_s(N)$
and  $r(N)$, in order to reconstruct the effective potential $V(\phi)$ 
and the non-minimal coupling function $F(\phi)$ as functions of the scalar field.

\begin{figure*}
\centering
\begin{subfigure}{0.45\textwidth}
\includegraphics[width=1\linewidth]{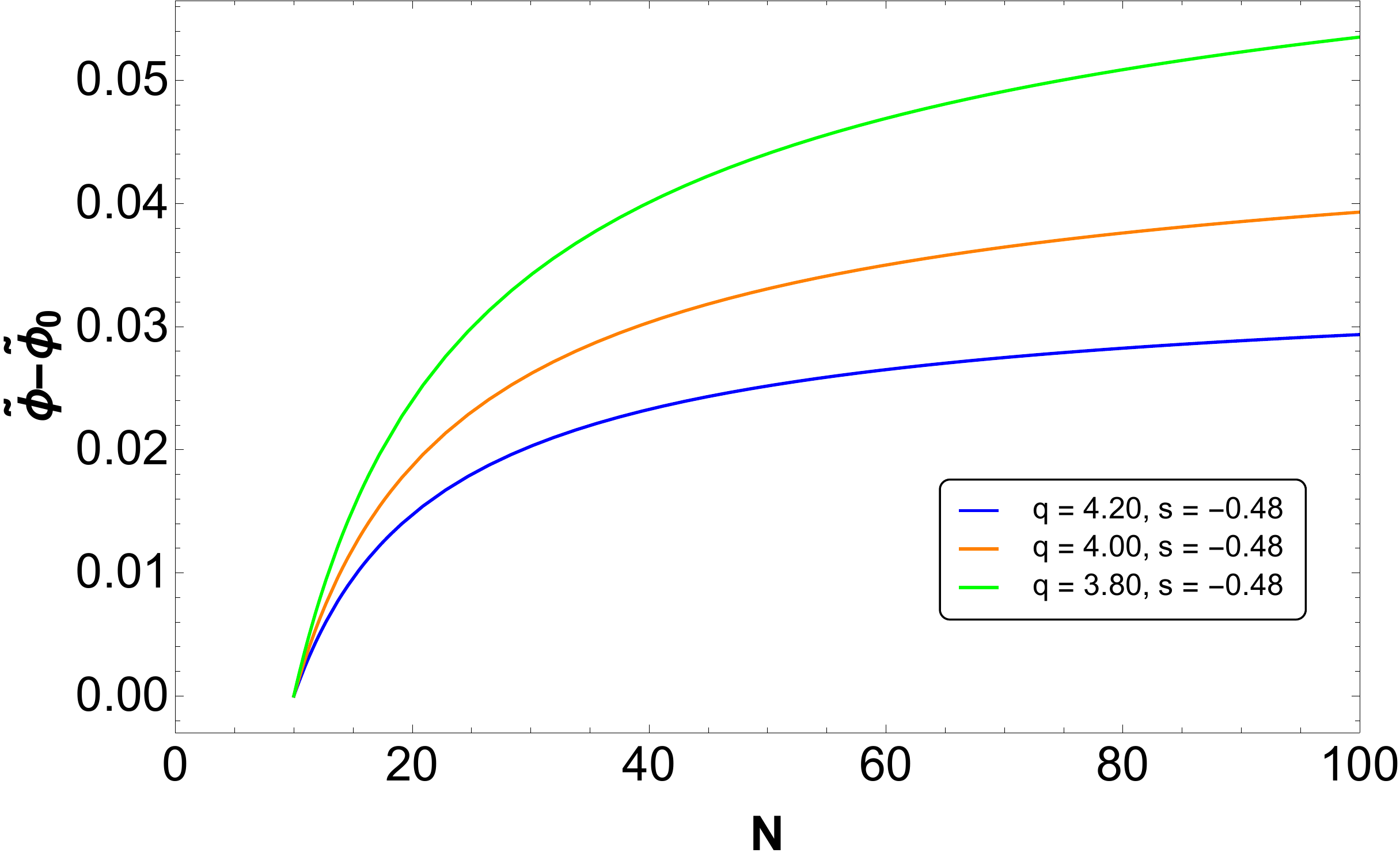} 
\end{subfigure}
\begin{subfigure}{0.45\textwidth}
\includegraphics[width=1\linewidth]{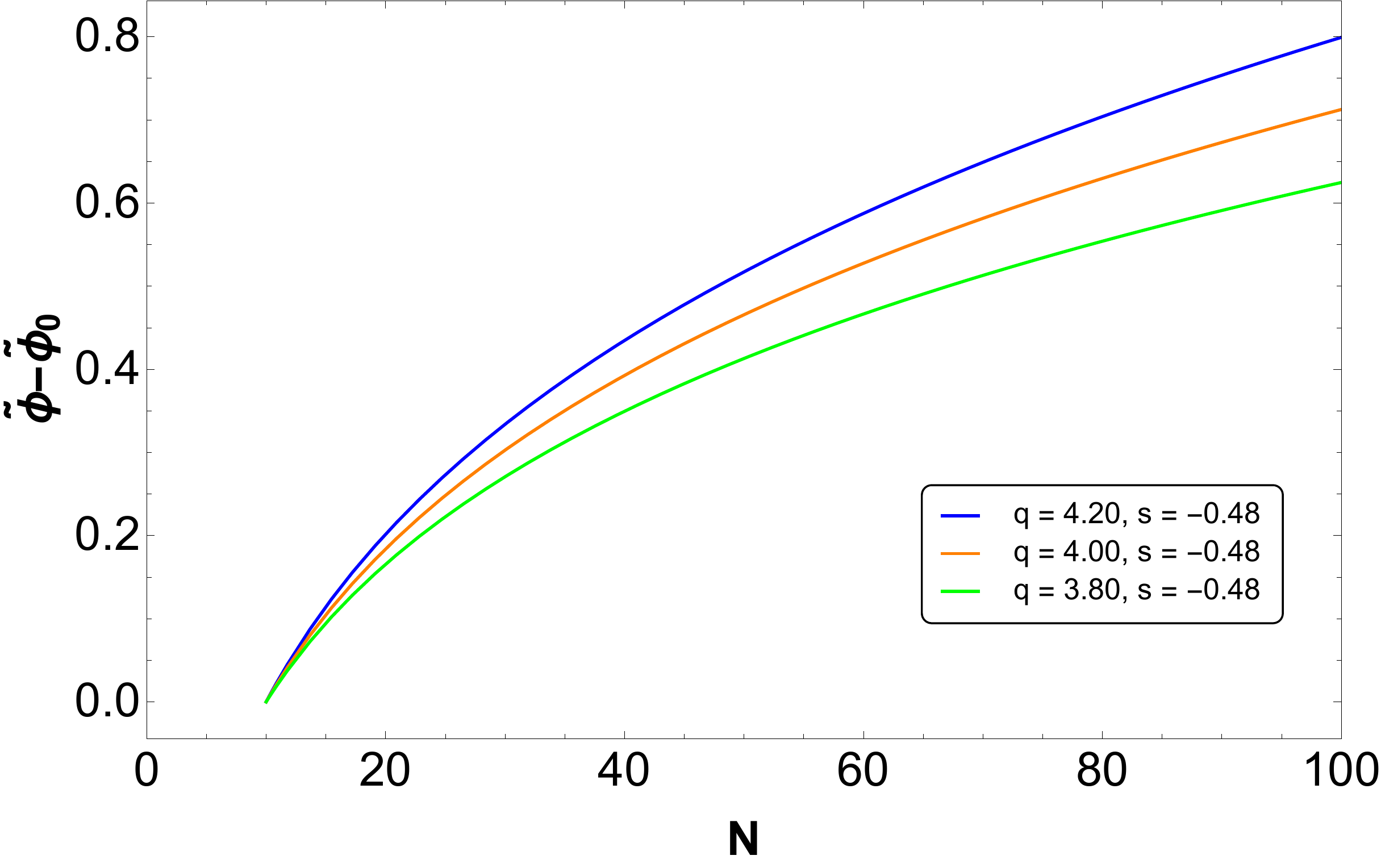}
\end{subfigure}
\begin{subfigure}{0.45\textwidth}
\includegraphics[width=1\linewidth]{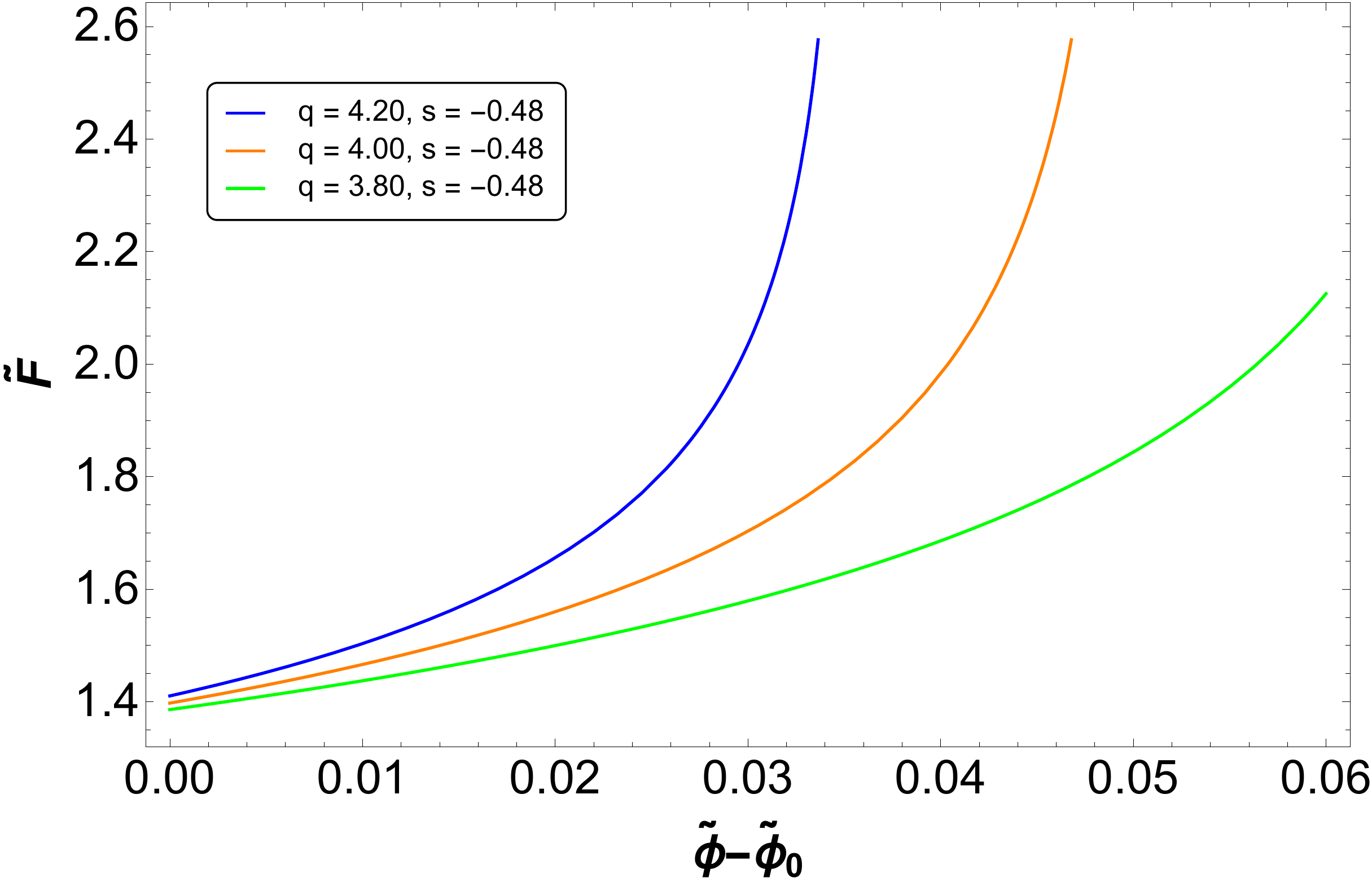} 
\end{subfigure}
\begin{subfigure}{0.45\textwidth}
\includegraphics[width=1\linewidth]{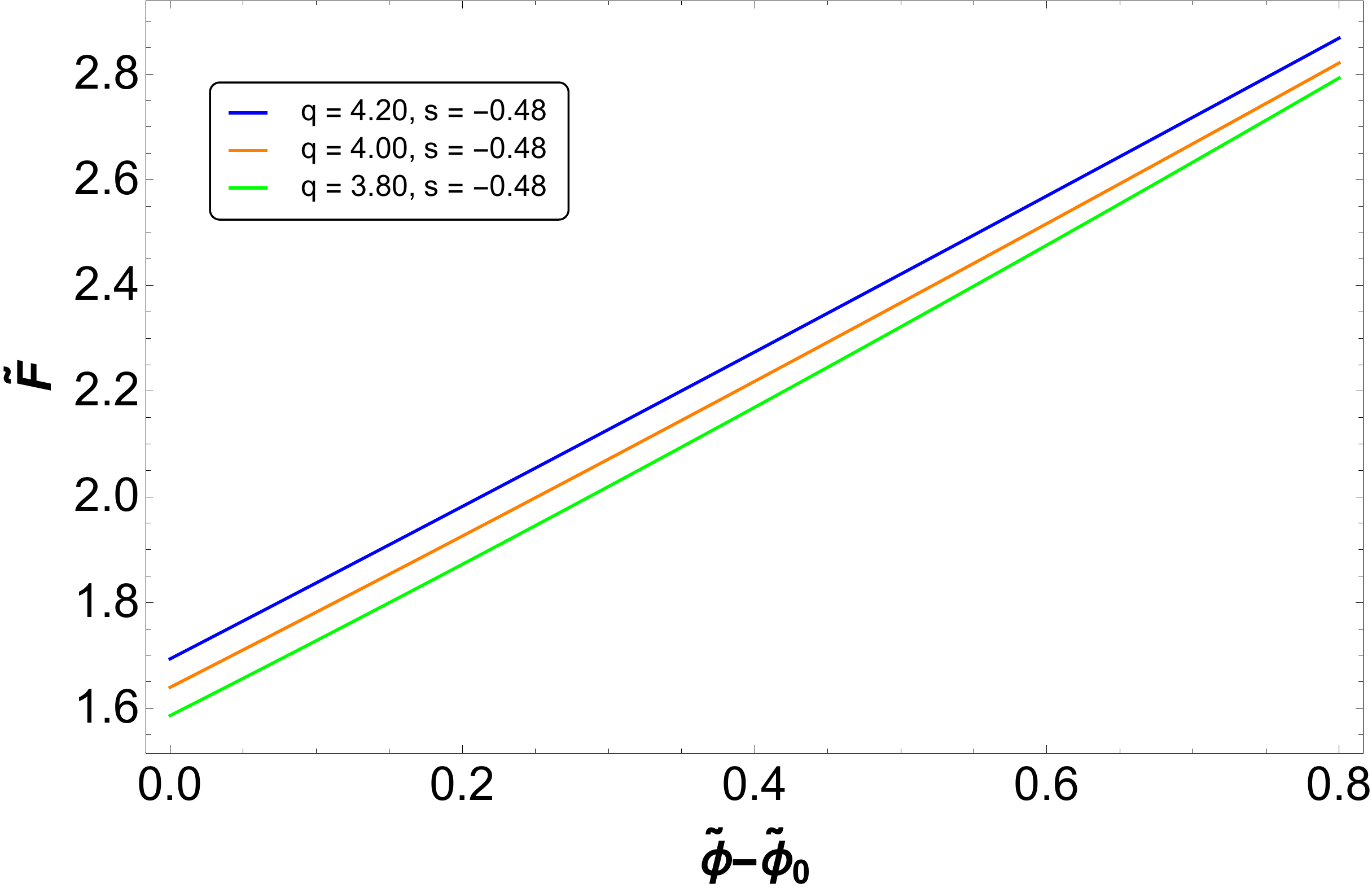}
\end{subfigure}
\begin{subfigure}{0.45\textwidth}
\includegraphics[width=1\linewidth]{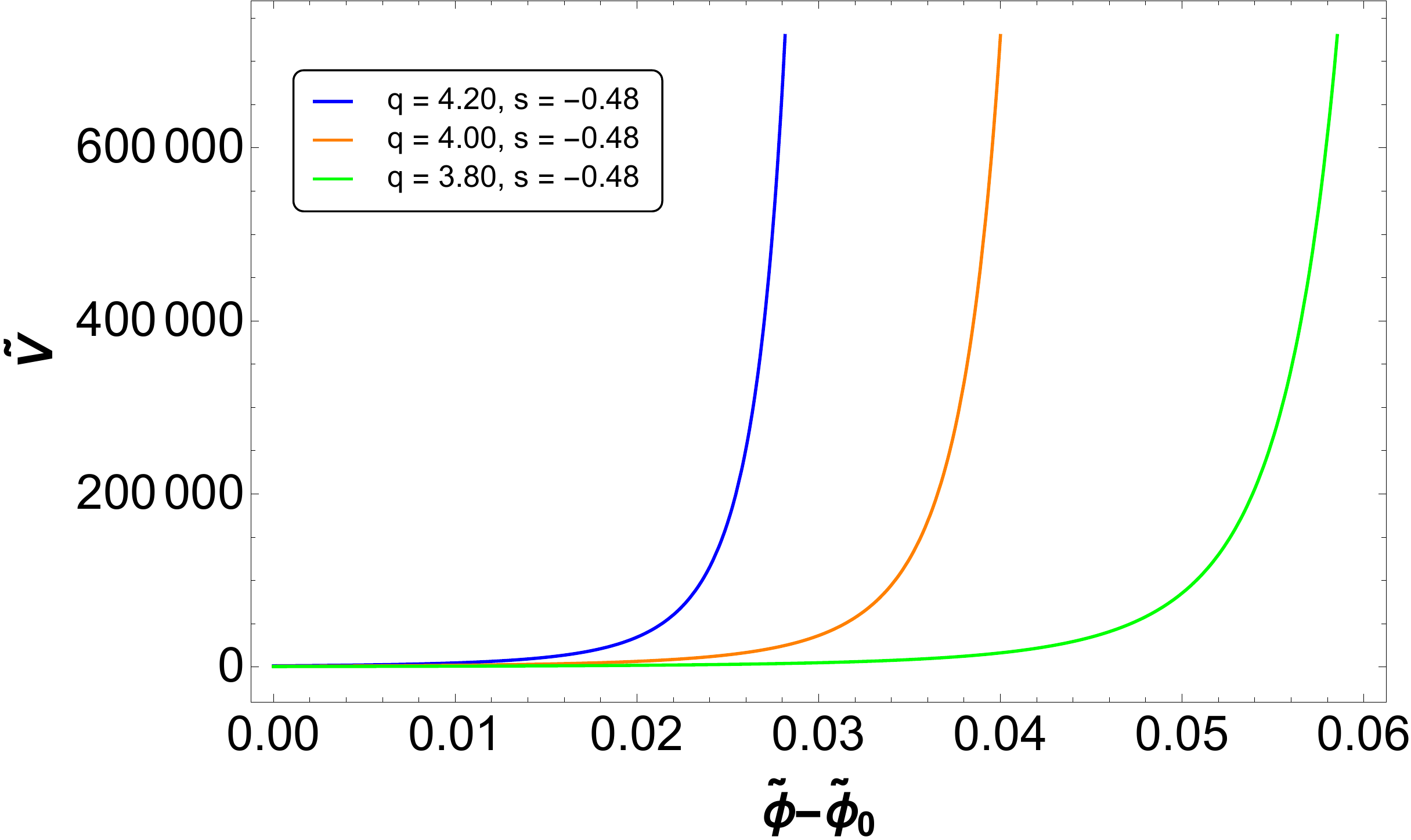} 
\end{subfigure}
\begin{subfigure}{0.45\textwidth}
\includegraphics[width=1\linewidth]{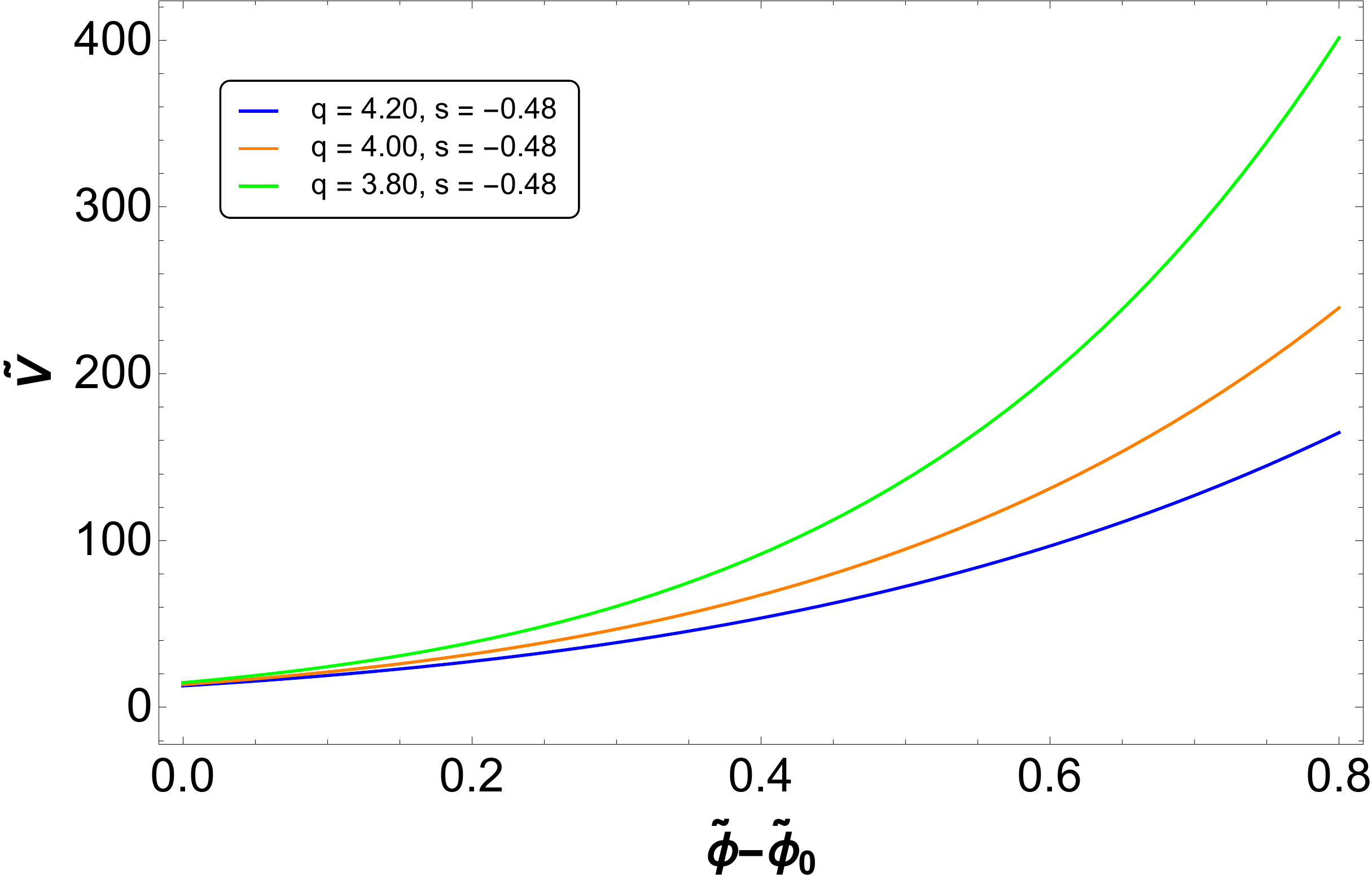}
\end{subfigure}
\caption{ It is shown the dependence of the scalar field $\tilde{\phi}-\tilde{\phi_0}$ versus the number of $e-$folds $N$ (upper panels). 
Also, we show the reconstruction of the redefined  non-minimal coupling function $\tilde{F}$ (center panels) and the  potential $\tilde{V}$
(lower panels) versus the dimensionless scalar field $\tilde{\phi}-\tilde{\phi}_0$. Here the left panels correspond to the first solution ($p_1$ and $p_2$ given by $+$ and $-$) and the right panels correspond to the second solution ($p_1$ and $p_2$ given by $-$ and $+$). In these plots we have used $N_0=10$.}
\label{fig1}
\end{figure*}

\subsection{Example 1}
To apply the methodology of above, we shall consider the simplest ansatz for large $N$ of  the observable parameters $n_s(N)$ and $r(N)$, in order to obtain the reconstruction of the background variables of our model; $V(\phi)$ and $F(\phi)$. Following, Ref. \cite{C} we can assume  that the scalar spectral index and the tensor-to-scalar ratio as functions of the number of $e-$folds are given by  
\bea
n_s(N)=1-\frac{2}{N},\:\:\:\:\mbox{and}\;\,\,\;\;\; r(N)=\frac{q}{N},
\label{attractors1}
\eea 
where $q$ is a dimensionless constant and it is positive.
 Note that  the parametrization for the spectral index  $n_s(N)$ as well  $r(N)$ are not allowed at point $N=0$ and these parametrizations are thought for large $N$. Here, large $N$ means that the number of $e-$folds $N$ corresponds to $N\sim\mathcal{O}(10)\sim\mathcal{O}(10^2)$ produced 
during the slow-roll scenario, see Refs. \cite{C,K}. We note that for the special case in which the number of $e-$folds $N=60$, the scalar spectral index is well corroborated by the Planck satellite. Also, in particular for the case in which the number  $N=60$ and the tensor-to-scalar ratio $r(N=60)<0.07$, we have that the upper bound for the  parameter $q$ is given by $q<4.2$

We observe that from Eq. (\ref{attractors1}) the relation between the tensor-to-scalar ratio and the scalar spectral index i.e., $r=r(n_s)$ or consistency relation results
$
r(n_{s})=q\left(1-n_{s}\right)/2.
$
Also, from relation (\ref{rcN}) and $q$ positive, we find  a lower bound for the parameter $q$ given by
\bea
q>\frac{64(1+2s)}{8+15s}>0,\,\,\,\,\,\,\mbox{if}\,\,\,\,\,\,0>s>-\frac{1}{2},\label{r1}
\eea
and the range 
\bea
\frac{64(1+2s)}{8+15s}>q>0,\,\,\,\,\,\,\mbox{if}\,\,\,\,\,\,s>0.\label{r2}
\eea
Here we note that Eq. (\ref{r2}) does not work, since the parameter $q<4.2$ from the observational data. Thus, from Eq. (\ref{r1}) we find that the range for the parameter $s$ in order to satisfy the constraint $q<4.2$ is given by
$-0.5<s<-0.468$.  In this way, from the ansatz given by relation (\ref{attractors1}),
we find that the parameter $s$ associated to the function $G(T)$ is negative, and it presents a narrow range given by $-0.5<s<-0.468$.

Now, from Eqs. \eqref{F_N_Strong2}, \eqref{V_N_Strong2}
and \eqref{attractors1}, we find that the non-minimal coupling function
$F$ and the effective potential $V$ in terms of the number of $e-$folds
are given by 
\bea
&& F(N)=M_{pl}^{2(1-s)}  N^{p_{1}} \xi,\\
&& V(N)=M_{pl}^4 N^{p_{2}} \lambda,
\eea where $\xi>0$, and $\lambda>0$ 
are two dimensionless integration constants and the powers $p_1$ and $p_2$ are 
 defined as
\bea
p_{1}&=&\frac{1}{32} s \left[-3 q\pm \Delta\right],\label{p1p20}\\
p_{2}&=&\frac{q (7 s+4)\mp s\; \Delta}{32 (2 s+1)},\label{p1p2}
\eea
respectively. Here the quantity $\Delta$ is given by $\Delta=\sqrt{q (-15 q s-8 q+128 s+64)/s}>0$ and the constant $s\neq 0$. Note that we have two branches of solutions for the functions $F(N)$ and $V(N)$ from the powers $p_1$
and $p_2$ defined in (\ref{p1p20}) and (\ref{p1p2}), respectively.  In this context, the first solution corresponds to the cases in which the powers $p_1$ and $p_2$ are given by the signs $+$ and $-$, in  Eq. (\ref{p1p20}) and Eq. (\ref{p1p2}), and the second  solution is associated to the signs $-$ and $+$ of $p_1$ and $p_2$, respectively.

Now, from Eq. \eqref{phi_N_Strong}, we find that the relation between the scalar field $\phi$ and  the number of $e-$folds $N$ can be written as
\bea
&& \frac{\phi -\phi_{0}}{M_{pl}}= \frac{1}{A} \Bigg[N^{\frac{1}{\sigma}}-N_0^{\frac{1}{\sigma}}\Bigg],\label{N123r}
\eea
and hence we can invert this  solution such that 
\bea
&& N(\phi)=\Bigg[N_{0}^{\frac{1}{\sigma}}+\frac{A (\phi -\phi_{0})}{M_{pl}} \Bigg]^{\sigma},\label{NNN1}
\eea 
where the constant  $\sigma$ is defined as 
\be
\sigma =\frac{2( 1+s)}{1+p_{1}+(1+p_{2})s},
\ee
and 
\be
A=\frac{1+p_{1}+(1+p_{2})s}{ q^\frac{1}{2} (s+1)^\frac{3}{2}(1+2 s)^{\frac{-s}{2(s+1)}} \;\lambda ^{\frac{s}{2(s+1)}} \xi ^{\frac{1}{2(s+1)}}}.
\ee
Besides, the number $N_0$ is defined as  $N(\phi=\phi_0)=N_0$ and we emphasize  that the range for the number of $e-$folds $N_0$ is given by  $0<N_0<N$. 
Additionally, in order to obtain a real and positive solution for the number of $e-$folds given by Eq. (\ref{NNN1}), we can consider  that the range for the effective scalar field $(\phi-\phi_0)$ becomes
\bea
\frac{\phi-\phi_0}{M_{pl}}>-\frac{N_0^{\frac{1}{\sigma}}}{A}\,,\,\,\,\,\mbox{if}\,\,\,\,\,1+p_{1}+(1+p_{2})s>0,
\eea
and 
\bea
\frac{\phi-\phi_0}{M_{pl}}<-\frac{N_0^{\frac{1}{\sigma}}}{A}\;,\,\,\,\,\mbox{if}\,\,\,\,\,1+p_{1}+(1+p_{2})s<0.
\eea

In this form, the reconstruction from the parametrization given by Eq. (\ref{attractors1}) for  the non-minimal coupling function and  the effective potential in terms of the scalar field $\phi$ can be written as 
\be
F(\phi)=\xi\,M_{pl}^{2(1-s)}\,\left[N_{0}^{\frac{1}{\sigma}}+\frac{A(\phi-\phi_{0})}{M_{pl}}\right]^{\sigma_1},\label{FE1}
\ee
and
\be
V(\phi)=\lambda\,M_{pl}^{4}\,\left[N_{0}^{\frac{1}{\sigma}}+\frac{A(\phi-\phi_{0})}{M_{pl}}\right]^{\sigma_2},\label{VE1}
\ee
where the constants $\sigma_1$ and $\sigma_2$ are defined as
$\sigma_1=\sigma p_1$ and $\sigma_2=\sigma p_2$, respectively.

 In particular we can assume  the condition
\be
\frac{A (\phi -\phi_{0})}{M_{pl} N_{0}^{\frac{1}{\sigma}}}\gg 1. \label{CE1}
\ee Under this condition, we can write
\be
N(\phi)=\tilde{N}_{0}(\phi -\phi_{0})^{\sigma},\label{NNN1_2}
\ee where we have defined
\be
\tilde{N}_{0}=\left[\frac{A}{M_{pl}}\right]^{\sigma}.
\ee

In this way, from Eq. (\ref{NNN1_2}), we find that the reconstruction for the non-minimal coupling function $F(\phi)$ and the effective potential considering the simplest  attractors given by (\ref{attractors1}) results
\be
F(\phi )= F_{0} (\phi -\phi_{0})^{ \sigma_1},
\label{F_Rec1}
\ee
and
\be
V(\phi )=V_{0} (\phi -\phi_{0})^{ \sigma_2 },
\label{V_Rec1}
\ee 
respectively. Here we have defined the constants $F_0$ and $V_0$ as
$F_{0}=M_{pl}^{2(1-s)} \tilde{N}_{0}^{p_{1}} \xi$ and $V_{0}=M_{pl}^{4}  \tilde{N}_{0}^{p_{2}} \lambda$, respectively. As we have considered that the velocity $\dot{\phi}<0$, then we can consider  the particular case in which $V_\phi>0$
and $F_{\phi}>0$ (see Eq. (\ref{dotphioverH}) for other combinations) with which the powers $\sigma_1$ and $\sigma_2$ must be positive.

In this sense, we find that  the reconstruction from  observable parameters $n_s(N)=1-2/N$ 
and $r=q/N$ gives rise to a chaotic potential and to a power-law for the 
non-minimal coupling function. As it is well known in the framework of the GR, the chaotic potential does not work because the tensor-to-scalar ratio is disapproved from the observational data. However, in our model based on the frame of the scalar- torsion $f(T,\phi)$ gravity,  we find that the chaotic model works but with the condition that the 
 function $F(\phi)$ is given by a power-law function.

In Fig.1 we show the evolution of the scalar field 
$\tilde{\phi}-\tilde{\phi_0}$ versus the number of $e-$folds $N$ (upper panels). Also, we show the 
reconstruction of the background variables $F(\phi)$ and $V(\phi)$
through   the redefined 
non-minimal coupling function $\tilde{F}$ (center panels) and the redefined  potential $\tilde{V}$ (lower panels)
versus the new field $\tilde{\phi}-\tilde{\phi_0}$, for the two branches of solutions, in order to satisfy the attractors  $n_s(N)$ and $r(N)$ given by (\ref{attractors1}). Here the left panels correspond to the first solution where the powers $p_1$ and $p_2$ are given by the signs $+$ and $-$, in Eq. (\ref{p1p20}) and Eq. (\ref{p1p2}), and the right panels are for the second  solution with the signs $-$ and $+$ of $p_1$ and $p_2$, respectively.
Besides, we have defined the dimensionless field $\tilde{\phi}-\tilde{\phi_0}$  as $\tilde{\phi}-\tilde{\phi_0}=(\phi-\phi_0)/(M_{pl}\lambda^{s/2(s+1)}\xi^{1/2(s+1)})$ and the dimensionless quantities $\tilde{F}$ and $\tilde{V}$ as $\tilde{F}=F/(\xi\,M_{pl}^{2(1-s)})$ and $\tilde{V}=V/(\lambda M_{pl}^4)$, respectively.

In order to write down the new field in terms of the number $N$ and the redefined functions $\tilde{F}$ and $\tilde{V}$
 as functions of the new field, we regard Eqs. (\ref{N123r}), (\ref{FE1}) and (\ref{VE1}) for different values of the parameter $q$ together with  the specific value of $s=-0.48$. In the panels we have considered that the number of $e-$folds $N_0=10$, see Ref. \cite{H}. From the upper panels, we observe that for both solutions  the scalar field tends to constant value for large $N$. From the center and lower panels, we note that  the reconstruction of the background variables follows a similar behavior to power law as described by Eqs. (\ref{F_Rec1}) and (\ref{V_Rec1}).

\subsection{Example 2}

As a second example to apply the methodology of above, we shall consider  other ansatz for large $N$ of  the observable parameter  $r(N)$, in order to obtain the reconstruction of the effective potential  $V(\phi)$ and the function $F(\phi)$. Following, Refs. \cite{C,Herrera:2018cgi,Herrera:2018mvo} we consider that the scalar spectral index and the tensor to scalar ratio in terms  of the number of $e-$folds become
\bea
n_s(N)=1-\frac{2}{N},\:\:\:\:\mbox{and}\;\,\,\;\;\; r(N)=\frac{q}{N (N+\gamma)},
\label{attractors}
\eea 
respectively. Here the quantities $q$ and $\gamma$ are two dimensionless constants. 
Thus, for $q>0$ we have that $N>-\gamma$ and for the case in which $q<0$ the parameter $\gamma<-N$,  such that the ratio $r>0$. In particular for the specific case in which $q=1$  different constraints  for the parameter $\gamma$ have been obtained in Refs. \cite{H,O}.
Also, the situation in which the parameter $q=12\sigma$ and $\gamma=0$ is called in the literature the $\sigma$ attractor and this model was studied in Refs. \cite{K,J}.
For the case of the T-model, we have that the tensor-to-scalar ratio corresponds to  $q=8/\sigma^2$ and $\gamma=0$, see Ref. \cite{T}.

From Eq. \eqref{attractors} we obtain that the consistency relation 
$r=r(n_{s})$ can be written as
\bea
&&r(n_s)=\frac{q}{2}\left[\frac{(1-n_s)^2}{[2+\gamma(1-n_s)]}\right].
\label{ns8}
\eea
Here we note that for the specific case in which the parameter $|\gamma|\gg (1-n_s)^{-1}$, we have that the consistency relation gives us $r(n_{s})\propto (1-n_s)$ and in the opposite limit
whenever $|\gamma|\ll (1-n_{s})^{-1}$, we get that $r(n_s)\propto (1-n_s)^2$.

Now, from Eqs. \eqref{F_N_Strong2}, \eqref{V_N_Strong2}
and \eqref{attractors}, we find that the coupling function and the effective potential in terms of the number of $e-$folds are given by 
\bea
\label{F_Ex1}
&& F(N)=M_{pl}^{2\left(1-s\right)}   \left[\frac{N}{\gamma +N}\right]^{p_{1}}\xi,\\
&& V(N)= M_{pl}^4 \left[\frac{N}{\gamma +N}\right]^{p_{2}} \lambda,
\label{F_Ex2}
\eea where the quantities $\xi$ and $\lambda$ correspond to two integration constants (dimensionless constants) different from zero. The powers $p_1$ and $p_2$ are defined as 
\bea
&& p_{1}= -\frac{s (3 q\pm\Delta)}{32 \gamma },\label{p12}\\
&& p_{2}= \frac{q (7 s+4)\mp s \Delta}{32 \gamma  (2 s+1)}\label{p22},
\eea 
where $\Delta=\sqrt{\frac{q (64 \gamma  (2 s+1)-q (15 s+8))}{s}}>0$  and $s\neq 0$. Here, recalled that the parameter $s>-1/2$. 
 As before we note that we have two branches of solutions for the functions $F(N)$ and $V(N)$.
 Also, assuming the cases in which the parameters $q$ and $\gamma$ are positive, we have
\bea
q>\frac{64\gamma(1+2s)}{8+15s}>0,\,\,\,\,\,\,\mbox{if}\,\,\,\,\,\,0>s>-\frac{1}{2},\label{r12}
\eea
and the range 
\bea
\frac{64\gamma(1+2s)}{8+15s}>q>0,\,\,\,\,\,\,\mbox{if}\,\,\,\,\,\,s>0.\label{r22}
\eea

To continue with the  reconstruction of the background variables, we need the relation between the number of $e-$folds and the scalar field i.e., $N=N(\phi)$.    
Thus, from \eqref{phi_N_Strong} we can write  
\be
\phi-\phi_{0}=\frac{M_{pl}}{2}\;\int_{N_0}^N \left[ \sqrt{\frac{q (s+1) \left(\frac{2 s+1}{\lambda}\right)^{\frac{1}{s+1}-1} \xi ^{\frac{1}{s+1}} }{N'^2 (1+\frac{\gamma}{N'})^{\frac{p_{1}+p_{2} s}{s+1}+1}}}\;\right]\,dN',
\label{phi_N_Eq}
\ee 
where again  $N_0$ denotes the number of $e-$folds at moment in which $N(\phi=\phi_0)=N_0$  and it satisfies $0<N_0<N$. From Eq. (\ref{phi_N_Eq}), we note that we cannot invest analytically the solution $N=N(\phi)$ to  generate the reconstruction. Here, we mention that  the integral on the number of $e-$ folds corresponds to a Hypergeometric function.

In this context and in order to obtain an  inverse solution for the relation $N=N(\phi)$, we can assume the
limit in which $\abs{\gamma}/N\ll 1$ (here the ratio $r\propto N^{-2}(1-\gamma/N+...)\sim N^{-2}$ as in Ref. \cite{L}).
The consistency relation $r=r(n_s)$ can be written as 
$r(n_s)\simeq (q/4)(1-n_s)^2[1-\gamma(1-n_s)/2+....]$.
In particular for the specific case in which $N=60$ and the tensor-to-scalar ratio $r<0.07$, we can estimate approximately   an upper bound for the parameter $q$ given by $252>q>0$.

Under the approximation in which $\abs{\gamma}/N\ll 1$, we can consider that the functions  $F(N)$ and $V(N)$ can be approximated to  $\mathcal{O}(\abs{\gamma}/N)$, such that 
$F(N)\simeq M_{pl}^{2\left(1-s\right)} \xi  \left[1-p_{1}\left(\frac{\gamma}{N}\right)+..\right]$, and $V(N)\simeq M_{pl}^{4} \lambda  \left[1-p_{2}\left(\frac{\gamma  }{N}\right)+..\right]$.  Hence, from equation \eqref{phi_N_Strong} we find that the solution $N=N(\phi)$ is given by
\be
N(\phi)\simeq N_0\,e^{\eta_1 (\phi -\phi_{0})},\label{NNN2}
\ee 
where the constant $\eta_1$ is defined as 
\be
\eta_1 =\frac{2}{M_{pl} \sqrt{q (s+1) \left(\frac{1+2 s}{\lambda}\right)^{\frac{1}{s+1}-1}\xi ^{\frac{1}{s+1}} }}.
\ee

In this form, considering  Eq. (\ref{NNN2}) we find that the reconstruction for the non-minimal coupling function and the effective potential in terms of the scalar field $\phi$ in the limit in which $\abs{\gamma}/N\ll 1$ becomes
\be
\label{Fphi_Ex1}
F(\phi)\simeq F_0\, \left[1-\frac{\gamma p_{1} e^{-\eta_{1}  (\phi -\phi_{0})}}{N_{0}}\right],
\ee and 
\be
V(\phi)\simeq V_0\, \left[1-\frac{\gamma p_{2} e^{-\eta_{1}  (\phi -\phi_{0})}}{N_{0}}\right],
\label{Vphi_Ex1}
\ee 
where $F_0=M_{pl}^{2\left(1-s\right)} \xi $
and $V_0=M_{pl}^{4} \lambda $, respectively. 
Here we note that the scalar potential \eqref{Vphi_Ex1} can be  similar to the Starobinsky potential if the parameter $\eta_{1}=\sqrt{2/3} \kappa$ \cite{Starobinsky:1980te}. In particular, we observe that  in the limit in which the quantity  $\eta_{1} (\phi -\phi_{0})\gg 1$, both the scalar potential and the non-minimal coupling become a constant.


\section{Concluding Remarks}\label{Concluding_Remarks}

In this article, we have studied the reconstruction of the inflationary epoch, 
in the framework of a general class of scalar-torsion theory, where the Lagrangian density is given by an arbitrary function $f(T,\phi)$, in which $T$ denotes the torsion scalar of teleparallel gravity and $\phi$ the inflaton field. In the context of the slow-roll approximation and   
under a general treatment  of reconstruction, we have obtained  expressions for the effective potential and the non-minimal coupling function in terms of the 
  cosmological parameters such as the scalar spectral index $n_s$ and the tensor-to-scalar ratio $r$.
 In this general analysis we have found from the parametrization of the cosmological quantities $n_s(N)$ and $r(N)$, in which the parameter  $N$ denotes  the number of $e-$folds, different relations  for the effective potential and non-minimal coupling in the high energy limit. For this energy limit in which $\mathcal{G}_{eff}/\mathcal{G}\ll 1$, with $\mathcal{G}_{eff}$ the effective gravitational constant, we have assumed a specific ansatz for the coupling function $G(T)$ given by  
 $G(T)=T^{s+1}$, with the power $s>-1/2$. In this respect, we mention that we cannot compare the full Lagrangian density $f(T,\phi)$ given by Eq. (\ref{Ayy}) with the observations from the attractors ($n_s$ and $r$),
 due to the difficulty to find analytical expressions for the background variables.
 However, in the particular case of the high energy limit, we  have been able to apply the reconstruction of these variables from the parametrization of the cosmological quantities $n_s(N)$ and $r(N)$. 
 
 Additionally, in order to explicate the reconstruction procedure for our model, we have taken into account that 
 we have three background variables to be reconstructed, the functions $G(T)$, $F(\phi)$ and $V(\phi)$, from  two cosmological quantities, $n_{s}(N)$ and $r(N)$. For simplicity and in order to find analytical solutions for the functions $F(\phi)$ and $V(\phi)$,  we have fixed the coupling function $G(T)$ and in particular we have considered the power-law form for this function.  Subsequently,  parameterizing $n_{s}$ and $r$ in terms of the number of $e$-folds $N$, 
we obtained a system of two decoupled equations (first-order) for the background variables $F(N)$ and $V(N)$.  Thus, from this decoupled system  
 we found the
general expressions for the non-minimal coupling function $F(N)$ and the effective scalar potential $V(N)$ in terms of the observables $n_{s}(N)$ and $r(N)$ together with the quantity
$dr(N)/dN$.
 
To apply the methodology  of  reconstruction  from the parametrization of the cosmological
observables $n_s(N)$ and $r(N)$ in the high energy regime
during the slow-roll approximation, we have used 
the simplest example of the scalar spectral index $n_s(N)$  
given by $n_s=1-2/N$ together with two examples associated to  the tensor-to-scalar ratio $r(N)$. In this context, we have assumed the specific cases in which the parametrization for the tensor-to-scalar ratio are given by  $r(N)\propto 1/N$ and
$r(N)\propto 1/(N[N+\gamma])$ with $\gamma=$ constant, in order to reconstruct the non-minimal coupling function $F(\phi)$ and the scalar potential $V(\phi)$.

For our first example,  in which the parametrization for the tensor-to-scalar ratio $r(N)\propto 1/N$, we have obtained that the reconstruction for the non-minimal coupling function $F(\phi)$ and the effective potential $V(\phi)$ evolves as power-law, see Eqs. (\ref{FE1}) and (\ref{VE1}). Here, we have noted that the reconstruction of these functions is not unique, since we  
have found two branches of solutions product  of the powers $p_1$ and $p_2$.  In the particular case given by  (\ref{CE1}), we have found that the non-minimal function and the effective potential are reduced to Eqs. 
(\ref{F_Rec1}) and (\ref{V_Rec1}), respectively. Also interestingly  in this example we have obtained that the chaotic potential works unlike  of the  GR.

In Fig. 1 we show the evolution of the scalar field 
$\tilde{\phi}-\tilde{\phi_0}$ versus the number of $e-$folds $N$ (upper panels). Also,  we show the 
reconstruction of the background variables $F(\phi)$ and $V(\phi)$
through   the redefined 
non-minimal coupling function $\tilde{F}$ (center panels) and the redefined  potential $\tilde{V}$ (lower panels)
versus the new field $\tilde{\phi}-\tilde{\phi_0}$, for the two branches of solutions.
Here we observed that for the both solutions  the scalar field tends to constant value for large $N$ (upper panels).  Also, 
 we noted that  the reconstruction of the background variables ($F(\phi)$ and $V(\phi)$) follows a similar behaviour to power law as described by Eqs. (\ref{F_Rec1}) and (\ref{V_Rec1}), see   center and lower panels.

For the second  example in which we give the observables  $n_s(N)$ and $r(N)$ given by Eq. (\ref{attractors}),  we have found    the non-minimal coupling function $F(N)$ and the potential $V(N)$ in terms of the number of $e-$folds, see  Eqs. (\ref{F_Ex1}) and (\ref{F_Ex2}). Here, as before, we
have obtained two branches of solutions for these functions given by the powers $p_1$ and $p_2$ defined by Eqs. (\ref{p12}) and (\ref{p22}), respectively.
 For this example we note that  we could not obtain an inverse  solution from Eq. (\ref{phi_N_Eq}) in order to find the number of $e-$folds in terms of the scalar field $\phi$ i.e., $N=N(\phi)$. In this way, we could not 
 rebuild the effective potential 
  $V(\phi)$ and the non-minimal coupling $F(\phi)$  analytically. 
 Thus, from Eqs. (\ref{F_Ex1}) and (\ref{F_Ex2})  we have studied  the
  potential and the coupling function in the limit in which the ratio $\abs{\gamma}/N\ll 1$, in order to find an analytical reconstruction of these background 
  variables.
  In this sense, under the approximation in which the ratio $\abs{\gamma}/N\ll$ 1, we have obtained that the dependence between the number $N$ and the scalar field $\phi$ changes exponentially, see relation (\ref{NNN2}). In this form, we have found that the  non-minimal coupling $F(\phi)$ and the effective potential $V(\phi)$ as a function of the scalar field $\phi$  are described by Eqs. (\ref{Fphi_Ex1}) and (\ref{Vphi_Ex1}), 
  respectively. Here we noted that for large values of the product $\eta_1(\phi-\phi_0)$, the coupling function $F(\phi)$ and the effective potential $V(\phi)$ tend to a constant value as the ultra slow-roll regime \cite{Martin:2012pe}.
  
Additionally, we can comment that after the end of inflation, there will be a period of reheating of the universe, characterized by a certain reheating temperature. Interestingly, it is possible to relate this temperature with the number $N$, considering
  that the reconstructed potential $V(N)$ is valid for small value of $e-$folds, see Refs. \cite{C,F1,F2}. Here a detailed analysis of the reheating scenario from the reconstructed potential (and coupling function) must be done in order to see how it affects the obtained results during the reconstruction of inflation. 

Finally, in the present article, we have obtained 
  the reconstruction of the non-minimal coupling function $F(\phi)$ and the scalar potential $V(\phi)$ fixing the function $G(T)$. 
In this context, the observational data from the Planck satellite and the parametrization of the cosmological observables $n_s(N)$ and $r(N)$ 
allow us to constrain the free parameters and to reconstruct the functions $F(\phi)$ and $V(\phi)$. In particular,  under the slow-roll approximation we  have found different potentials, and their corresponding associated non-minimal coupling functions. Consequently, in this context, our inflationary model is consistent with observations and, therefore, 
it is an alternative to general relativity but searching for a different viable condition is beyond the scope of this investigation work. Also, we have not addressed the reconstruction of the background variables 
 for other types of coupling functions $G(T)$ and other attractors $n_s(N)$ and $r(N)$.  We hope to return to these points in the near future.

\section*{Acknowledgments}
M. Gonzalez-Espinoza acknowledges support from PUCV. G. Otalora acknowldeges DI-VRIEA for financial support through Proyecto Postdoctorado $2020$ VRIEA-PUCV.





\end{document}